\documentclass[twocolumn,showpacs,showkeys,preprintnumbers,superscriptaddress,amsmath,floatfix,amssymb,secnumarabic]{revtex4}
\usepackage[colorlinks=true]{hyperref}
\usepackage{graphicx}
\usepackage{epsfig}

\def\({\left(}
\def\){\right)}
\def\[{\left[}
\def\]{\right]}

\newcommand{\lr}[1]{ \left( #1 \right) }
\newcommand{\lrs}[1]{ \left[ #1 \right] }

\newcommand{\vev}[1]{ \langle \, #1 \, \rangle }
\newcommand{\diag}[1]{ {\rm diag} \, \left( #1 \right) }

\newcommand{\expa}[1]{ \exp{\left( #1 \right)} }

\newcommand{\comment}[1]{}

\newcommand{\logo}{\\ \vskip -18mm
\leftline{\includegraphics[scale=0.3,clip=false]{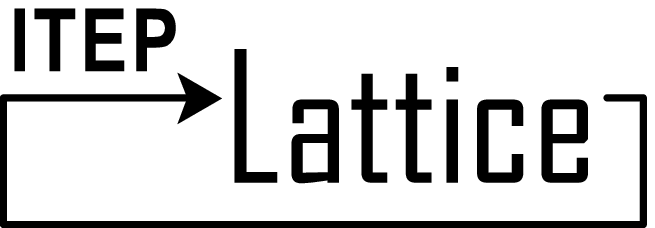}} \vskip 10mm}

\begin{document}
\sloppy

\title{Numerical study of the conductivity of graphene monolayer within the effective field theory approach  \logo}

\author{P.~V.~Buividovich}
\email{buividovich@itep.ru}
\affiliation{ITEP, B. Cheremushkinskaya str. 25, Moscow, 117218 Russia}
\affiliation{JINR, Joliot-Curie str. 6, Dubna, Moscow region, 141980 Russia}

\author{E.~V.~Luschevskaya}
\email{luschevskaya@itep.ru}
\affiliation{ITEP, B. Cheremushkinskaya str. 25, Moscow, 117218 Russia}

\author{O.~V.~Pavlovsky}
\email{ovp@goa.bog.msu.ru}
\affiliation{Institute for Theoretical Problems of Microphysics, Moscow State University, Moscow, 119899 Russia}
\affiliation{ITEP, B. Cheremushkinskaya str. 25, Moscow, 117218 Russia}

\author{M.~I.~Polikarpov}
\email{polykarp@itep.ru}
\affiliation{ITEP, B. Cheremushkinskaya str. 25, Moscow, 117218 Russia}

\author{M.~V.~Ulybyshev}
\email{ulybyshev@goa.bog.msu.ru}
\affiliation{Institute for Theoretical Problems of Microphysics, Moscow State University, Moscow, 119899 Russia}
\affiliation{ITEP, B. Cheremushkinskaya str. 25, Moscow, 117218 Russia}

\date{June 20, 2012}

\begin{abstract}
 We report on the direct numerical measurements of the conductivity of graphene monolayer. Our numerical simulations are performed in the effective lattice field theory with noncompact $3 + 1$-dimensional Abelian lattice gauge fields and $2 + 1$-dimensional staggered lattice fermions. The conductivity is obtained from the Green-Kubo relations using the Maximum Entropy Method. We find that in a phase with spontaneously broken sublattice symmetry the conductivity rapidly decreases. For the largest value of the coupling constant used in our simulations $g = 4.5$, the DC conductivity is less than the DC conductivity in the weak-coupling phase (at $g < 3.5$) by at least three orders of magnitude.
\end{abstract}

\maketitle

\section{Introduction}
\label{IntroductionSec}

 Graphene, a single layer of carbon atoms which form a two-dimensional honeycomb lattice, is probably the most widely discussed material in modern condensed
matter physics. A peculiar feature of charge carriers in graphene is that their energy spectrum near the Fermi point is similar to that of the free
 $2+1$-dimensional massless Dirac fermions. This explains the unusual transport properties of graphene, such as Klein tunneling or novel types of the quantum Hall effect
\cite{Novoselov:04:1, Novoselov:09:1, Geim:07:1}.

 The four spinor components of these Dirac fermions correspond to charge carriers which are localized on one of the two elementary rhombic sublattices of
  the honeycomb lattice and which are close to one of the two distinct Fermi points in the Brillouin zone of graphene. In this low-energy description two
  components of the non-relativistic spin are treated as two independent fermionic flavors. Interactions between fermions are mediated by electromagnetic
   fields which propagate freely in $3+1$-dimensional space. The strength of electromagnetic interactions can be controlled by placing graphene
    layers on substrates with different dielectric permittivities.

 Since charge carriers in graphene propagate with speed $v_F \approx c/300$, the effective coupling constant for electromagnetic interactions turns out to be quite large, $\alpha = \alpha_0/v_F = 300/137 \approx 2$ for suspended graphene \cite{Novoselov:04:1, Novoselov:09:1, Geim:07:1}. In this case the non-perturbative effects could play an important role. Theoretical considerations suggest that such strong interaction between charge carriers could result in the insulator-semimetal phase transition in graphene \cite{Leal:04:1, Gamayun:10:1, Araki:10:1, Araki:10:2, Araki:12:1, Son:07:1}.

 However, due to the large value of the effective coupling constant there are no reliable analytical methods which allow to study this phase transition from the first principles, and one has to use numerical simulations. The effective field theory of graphene can be efficiently simulated using lattice staggered fermions \cite{Lahde:09:1, Lahde:09:2, Lahde:09:3, Drut:10:1, Lahde:11:1, Hands:08:1, Hands:10:1, Hands:11:1}. A single flavor of staggered fermions on $2 + 1$-dimensional square lattice corresponds to two independent flavors of continuum Dirac fermions \cite{Burden:87:1, MontvayMuenster, DeGrandDeTarLQCD}, which exactly reproduces the number of fermion flavors in the graphene effective field theory.

 In papers \cite{Lahde:09:1, Lahde:09:2, Lahde:09:3, Drut:10:1, Lahde:11:1, Hands:08:1, Hands:10:1, Hands:11:1} the insulator-semimetal phase transition was studied numerically by considering the fermionic``chiral'' condensate $\langle \bar{\psi} \psi \rangle$. Within the effective field theory of Dirac quasiparticles, nonzero condensate signals the opening of a gap in the quasiparticle spectrum, thus it plays the role of the order parameter for the semimetal-insulator phase transition. In \cite{Lahde:09:1, Lahde:09:2, Lahde:09:3, Drut:10:1, Lahde:11:1} Coulomb interactions between fermions were modeled  by a non-compact $3 + 1$-dimensional Abelian lattice gauge field. It was found that the condensate is formed at a critical coupling constant of the non-compact gauge field $\beta \sim 0.1$. Motivated by the theoretical considerations of \cite{Son:07:1, Shankar:94:1}, the authors of \cite{Hands:08:1, Hands:10:1, Hands:11:1} have also studied a similar theory with a contact interaction instead of Coulomb potential. They have also found that fermionic condensate is formed at sufficiently strong coupling.

 However, the value of the conductivity has not yet been directly measured in numerical simulations. In this paper we report on direct numerical measurements of AC and DC conductivities of graphene within the effective field theory of Dirac quasiparticles. Our lattice regularization of this effective field theory is similar to the one used in \cite{Lahde:09:1, Lahde:09:2, Lahde:09:3, Drut:10:1, Lahde:11:1}. The conductivity is obtained from the Green-Kubo dispersion relations for the ground-state correlators of electromagnetic currents. These relations are inverted with the help of the Maximum Entropy Method \cite{Asakawa:01:1, Aarts:07:1}.

 In agreement with the predictions of \cite{Leal:04:1, Gamayun:10:1, Araki:10:1, Araki:10:2, Araki:12:1, Son:07:1, Lahde:09:1, Lahde:09:2, Lahde:09:3, Drut:10:1, Lahde:11:1, Hands:08:1, Hands:10:1, Hands:11:1} we find that when a nonzero fermionic condensate is formed, the DC conductivity rapidly decreases. For the maximal value of the coupling constant used in our simulations, which corresponds to substrate dielectric permittivity $\epsilon = 1.75$, the DC conductivity is smaller than the DC conductivity in the weak-coupling limit by at least three orders of magnitude.

 The paper is organized as follows: in Section \ref{sec:graphene_eft} we briefly review the effective field theory of graphene and discuss its lattice regularization, as well as suitable simulation algorithms. In Section \ref{sec:num_results} we present and discuss our numerical results for the graphene conductivity and the fermionic condensate $\vev{\bar{\psi} \psi}$. Section \ref{sec:conclusions} contains some concluding remarks and the discussion of the obtained results.

\section{Lattice regularization of the effective field theory of graphene}
\label{sec:graphene_eft}

\subsection{Basic definitions}
\label{subsec:basic_defs}

 We start from the Euclidean path integral representation of the partition function of the effective field theory of graphene \cite{Novoselov:04:1, Novoselov:09:1, Geim:07:1}:
\begin{eqnarray}
 \mathcal{Z}  =
 \int \mathcal{D}\bar{\psi}_f\mathcal{D}\psi_f \,
 \mathcal{D}A_{\mu} \,
 \exp \left(
 - \frac{1}{2}\int d^4x \,
 \lr{\partial_{\left[\mu\right.} A_{\left.\nu\right]}}^2
 - \right. \nonumber \\ -
 \int d^3x \,
 \bar{\psi}_f \, \Gamma_{0} \, \lr{\partial_{0} - i e A_{0}} \, \psi_f
 - \nonumber \\ \left. -
 \sum\limits_{i=1,2} \int d^3x \,
 \bar{\psi}_f \, \Gamma_{i} \, \lr{\partial_{i} - i e \, v_F \, A_{i}} \, \psi_f
 \right)  , \label{init_action}
\end{eqnarray}
where $A_{\mu}$, $\mu = 0 \ldots 3$ is the vector potential of the $3 + 1$ electromagnetic field, $\Gamma_{\mu}$ are Euclidean gamma-matrices and $\psi_f$, $f = 1, 2$ are two flavours of Dirac fermions which correspond to the two spin components of the non-relativistic electrons in graphene. We have also taken into account that Dirac fermions propagate in $2 + 1$-dimensional subspace at $x^3 = 0$ with speed $v_F \approx 1/300$.

 After rescaling of the coordinates and the vector potential
\begin{eqnarray}
\label{rescaling}
 x^0 \rightarrow x^0/v_F,
 \quad
 A^0 \rightarrow \sqrt{v_F} \, A^0,
 \quad
 A_i \rightarrow \frac{1}{\sqrt{v_F}} \, A_i \quad .
\end{eqnarray}
we conclude that the fluctuations of the spatial components $A_i$ of the vector potential are suppressed by a factor $1/v_F$ and we can set $A_i = 0$ in practical calculations. We thus arrive at the following partition function:
\begin{eqnarray}
\label{partfan}
 \mathcal{Z}   =
 \int \mathcal{D}\bar{\psi} \mathcal{D}\psi \mathcal{D}A_0
 \exp\left( -\frac{1}{2}\int d^4x \lr{\partial_{i} A_{0}}^2
 - \right. \nonumber\\ \left. -
 \int d^3x \, \bar{\psi}_f \, \lr{ \Gamma_0 \, \lr{\partial_0 - i g A_0}
 - \sum\limits_{i=1,2}
    \Gamma_i \partial_i} \psi_f
 \right) ,
\end{eqnarray}
 where the effective coupling constant $g^2 = e^2/v_F \approx 300/137 \sim 2$. Finite temperature $T$ can be introduced by imposing periodic boundary conditions in Euclidean time $x^0$ with the period $\frac{v_F}{k T}$.

 By virtue of the commutation relations $[O_a,\Gamma_0 \Gamma_i] = 0$ with $O_a = 1, \Gamma^3, \, \Gamma^5, \, i \Gamma^3 \Gamma^5$ the action of the effective field theory (\ref{partfan}) has the global $U(4)$ symmetry
\begin{eqnarray}
\label{symmetry}
\psi_f \rightarrow \expa{ i O_a \otimes \tau_b \, \alpha^{ab} } \, \psi_f  ,
\end{eqnarray}
where $\tau_0 = 1$ and $\tau_i$ - are spin Pauli matrices which act on flavor index $f$. Finite temperature and chemical potential do not break this global symmetry on the level of the Lagrangian, however, it might be broken spontaneously due to sufficiently strong Coulomb interactions \cite{Leal:04:1, Gamayun:10:1, Araki:10:1, Araki:10:2, Araki:12:1, Son:07:1, Lahde:09:1, Lahde:09:2, Lahde:09:3, Drut:10:1, Lahde:11:1, Hands:08:1, Hands:10:1, Hands:11:1}.

\subsection{Lattice action}
\label{subsec:latt_action}

 Following \cite{Lahde:09:1, Lahde:09:2, Lahde:09:3, Drut:10:1, Lahde:11:1, Hands:08:1, Hands:10:1, Hands:11:1} we use staggered fermions \cite{MontvayMuenster, DeGrandDeTarLQCD}
  in order to discretize the fermionic part of the action in (\ref{partfan}). One flavor of staggered fermions in $2 + 1$
   dimensions corresponds to two flavors of continuum Dirac fermions \cite{Burden:87:1, MontvayMuenster, DeGrandDeTarLQCD},
   which makes them especially suitable for simulations of the graphene effective field theory.

 The action for staggered fermions coupled to Abelian lattice gauge field is
\begin{eqnarray}
\label{lat_act_interact}
S_{\Psi}\lrs{\bar{\Psi}_x, \Psi_x, \theta_{x, \, \mu}} =
 \sum\limits_{x, y} \bar{\Psi}_x \, D_{x, y}\lrs{\theta_{x, \, \mu}} \, \Psi_y
= \nonumber \\ =
 \frac{1}{2} \, \sum\limits_{x} \, \delta_{x_3, \, 0} \, \left(  \sum\limits_{\mu=0, 1, 2}
 \bar{\Psi}_x \alpha_{x, \mu} e^{i \theta_{x, \, \mu}} \Psi_{x+\hat{\mu}}
 - \right. \nonumber \\ \left. -
 \sum_{\mu=0, 1, 2}
 \bar{\Psi}_x \alpha_{x, \mu} e^{-i \theta_{x, \, \mu}} \Psi_{x-\hat{\mu}}
 + m {\bar{\Psi}}_x \Psi_x \right) ,
\end{eqnarray}
where the lattice coordinates $x$ take integer values $x^{\mu} = 0 \ldots L_{\mu}-1$ and $x^3$ is restricted to $x^3 = 0$,
 $\bar{\Psi}_x$ is a single-component Grassman-valued field, $\alpha_{x, \mu} = (-1)^{x_0 + \ldots + x_{\mu-1}}$, and $\theta_{x, \, \mu}$ are the link variables which are the lattice counterpart of the vector potential $A_{\mu}\lr{x}$. For further convenience, we have also introduced the matrix elements $D_{x, y}$ of the staggered Dirac operator. The fields $\bar{\Psi}_x$, $\Psi_x$ satisfy periodic boundary conditions in spatial directions and anti-periodic boundary conditions in the Euclidean time direction. To account for the latter, we make a shift $\theta_{x, 0} \rightarrow \theta_{x, 0} + \pi$ in (\ref{lat_act_interact}) at the time slice with $x^0 = 0$.

 In order to recover the original spinor and flavor indices of the continuum Dirac fermions in (\ref{partfan}), the lattice should be subdivided into the cubic blocks consisting of $2 \times 2 \times 2$ elementary lattice cells. Now the coordinates of all lattice sites can be enumerated as $x_{\mu} = 2 y_{\mu} + \eta_{\mu}$, where $\eta_{\mu} = 0, 1$. We define the new fields on the lattice of $y$ coordinates \cite{MontvayMuenster, DeGrandDeTarLQCD, Lahde:09:2}:
\begin{eqnarray}
\label{staggered_org}
 \lrs{\Phi_y}^{\alpha}_f = \frac{1}{4 \sqrt{2}} \sum_{\eta}
 \lrs{\Gamma_0^{\eta_0} \Gamma_1^{\eta_1} \Gamma_2^{\eta_2} }^{\alpha}_f
 W_{y, \, \eta}
 \Psi_{2 y + \eta}  ,
\end{eqnarray}
where $W_{y, \, \eta}$ is the product of $e^{i \theta_{x, \mu}}$ along the the path which connects lattice sites with coordinates $2 y$ and $2 y + \eta$, $\alpha = 1, 2, 3, 4$ is the Dirac spinor index and $f = 1, ..., 4$ is the flavor index. It can be shown that in terms of these new fields defined on the lattice with double lattice spacing the staggered fermion action (\ref{lat_act_interact}) reproduces the naive discretization of the continuum fermionic action in (\ref{partfan}). However, there are additional terms which explicitly break the global $U\lr{4}$ symmetry of the continuum action in (\ref{partfan}) down to its $U\lr{1} \otimes U\lr{1}$ subgroup and which decouple only in the long-wavelength limit \cite{MontvayMuenster, DeGrandDeTarLQCD, Lahde:09:2}.

 Since the fermion action (\ref{lat_act_interact}) is restricted to the $2 + 1$-dimensional subspace with $x^3 = 0$, not all components of $\lrs{\Phi_y}^{\alpha}_f$ are independent. Due to the absence of $\Gamma^3$ in (\ref{staggered_org}) it satisfies the constraint
\begin{eqnarray}
\Gamma_3 \, \Gamma_5 \, \Phi_y \, \Gamma_5 \, \Gamma_3  = \Phi_y  .
\end{eqnarray}
It is easy to check that in a representation of Euclidean gamma-matrices with $\Gamma_3 \, \Gamma_5 = \diag{1, 1, -1, -1}$ this constraint implies the following block-diagonal form of the matrices $\Phi_y$:
\begin{eqnarray}
\Phi = \left(\begin{array}{cc}
 A & 0\\
 0 & B
 \end{array}\right)  ,
\end{eqnarray}
which is equivalent to two flavors of $4$-component Dirac spinors.

 Now let us consider lattice discretization of the action of the electromagnetic field in (\ref{partfan}).
  There exist two basic formulations of the $U\lr{1}$ lattice gauge theory: compact and non-compact.
   In order to exclude non-physical confining phase of the compact $U\lr{1}$ gauge theory \cite{PolyakovGaugeStrings} here we use the non-compact action for the gauge fields:
\begin{eqnarray}
\label{gauge_lat_act}
 S_g\lrs{\theta_{x, \, \mu}} = \frac{\beta}{2} \, \sum\limits_x \sum\limits^{3}_{i=1}
 \lr{ \theta_{x, \, 0} - \theta_{x + \hat{i}, \, 0} }^2  ,
\end{eqnarray}
where summation over $x$ now goes over the whole four-dimensional lattice. As discussed above, the fluctuations of the spatial components of the vector potential $A_i\lr{x}$
are suppressed in the effective field theory of graphene (\ref{partfan}). Correspondingly, we also set to zero the spatial link variables $\theta_{x, \, i}$.

 In continuous space, the inverse lattice coupling constant $\beta$ is related to the substrate dielectric permittivity $\epsilon$ as
\begin{eqnarray}
\label{lattice_coupling_constant}
 \beta \equiv \frac{1}{g^2} = \frac{v_F}{4 \pi e^2} \, \frac{\epsilon + 1}{2} ,
\end{eqnarray}
where the factor $\frac{\epsilon + 1}{2}$ takes into account the screening of the electrostatic interactions by the substrate. However, this relation can be modified due to finite lattice spacing effects such as the flavor symmetry breaking for staggered fermions \cite{Giedt:11:1, Lahde:10:1}. Generally, such effects tend to shift the phase transition towards the weak-coupling region \cite{Giedt:11:1, Lahde:10:1}. We leave the study of such finite-spacing artifacts for future work.

 We note also that although the gauge field action (\ref{gauge_lat_act}) is non-compact, the fermionic action (\ref{lat_act_interact}) is still ``compact'', that is, periodic in the variables $\theta_{\mu}\lr{x}$. In general, it is impossible to couple the gauge field to lattice fermions in a non-compact way while preserving the gauge invariance of the theory.

 Since the fermionic action (\ref{lat_act_interact}) is bilinear in the fermion fields, they can be integrated out in the partition function (\ref{partfan}):
\begin{eqnarray}
\label{ferm_integrated}
 \mathcal{Z} =
 \int \mathcal{D}\bar{\Psi}_x \, \mathcal{D}\Psi_x \, \mathcal{D}\theta_{x, \, 0}\,
 \nonumber \\
 \exp\lr{ - S_g\lrs{\theta_{x, \, 0}} - S_{\Psi}\lrs{\bar{\Psi}_x, \Psi_x, \theta_{x, \, 0}}}
 = \nonumber \\ =
 \int \mathcal{D}\theta_{x, \, 0}\,
 \det\lr{ D\lrs{ \theta_{x, \, 0} } }
 \expa{ -S_g\lrs{\theta_{x, \, 0}} }  .
\end{eqnarray}
Thus we deal with the effective action
\begin{eqnarray}
\label{eff_action}
 S_{eff}\lrs{ \theta_{x, \, 0} } = S_g\lrs{ \theta_{x, \, 0} } - \ln \det\lr{ D\lrs{ \theta_{x, \, 0} } }
\end{eqnarray}
which includes the determinant $\det\lr{ D\lrs{ \theta_{x, \, \mu} }}$ of the staggered Dirac operator $D_{x, y}\lrs{\theta_{x, \, \mu}}$ introduced in (\ref{lat_act_interact}).

\subsection{Simulation algorithm}
\label{subsec:algorithms}

 We use the standard Hybrid Monte-Carlo Method for generation of the configurations of the field $\theta_{x, \, 0}$  with the statistical weight $\expa{-S_{eff}\lrs{ \theta_{x, \, 0} }}$
 \cite{MontvayMuenster, DeGrandDeTarLQCD, Lahde:09:2}.

 In order to calculate the determinant of the staggered Dirac operator in (\ref{ferm_integrated}), we take into account that in the basis of even and odd lattice sites (which are defined as lattice sites with even or odd sum of all coordinates $x^0 + x^1 + x^2$) it takes the form \cite{MontvayMuenster, DeGrandDeTarLQCD, Lahde:09:2}
\begin{eqnarray}
\label{det_str}
 D\lrs{ \theta_{x, \, 0} } = \left(
  \begin{matrix}
    m & D_{eo} \\
    D_{oe} & m
  \end{matrix}
 \right)
\end{eqnarray}
with $D_{eo}^{\dag} = - D_{oe}$. The determinant of $D$ is thus equal to
\begin{eqnarray}
\label{det_fin}
\det\lr{D} = \det\lr{m^2 + D^{\dag}_{eo} \, D_{eo}}  ,
\end{eqnarray}
which is a manifestly positive quantity. Note that the operator $m^2 + D^{\dag}_{eo} \, D_{eo}$ acts only on the subspace of even lattice sites. We use the $\Phi$-algorithm
in our simulations \cite{MontvayMuenster, DeGrandDeTarLQCD}, in which the determinant (\ref{det_fin}) is represented in terms of a Gaussian integral over the pseudo-fermion field $\phi_x$:
\begin{eqnarray}
\label{pseudoferm_distr}
 \det\lr{m^2 + D^{\dag}_{eo} \, D_{eo} }
 = \int \mathcal{D}\bar{\phi}_x \, \mathcal{D}\phi_x
 \nonumber \\
 \expa{ - \sum\limits_{x, y} \bar{\phi}_x \,
 \lr{m^2 + D^{\dag}_{eo} \, D_{eo}}^{-1}_{x, y}
  \, \phi_y }  ,
\end{eqnarray}
where the sum over $x, y$ goes only over even lattice sites. The field $\phi_x$ is then stochastically sampled with the weight (\ref{pseudoferm_distr}). To this end we generate
 the random field $\xi_x$ according to the Gaussian distribution $P\lrs{\xi_x} \sim \expa{ - \sum\limits_{x} \bar{\xi}_x \, \xi_x }$ and then calculate
  $\phi_{x} = \sum\limits_{y} \lr{m^2 + D^{\dag}_{eo} \, D_{eo}}^{-1}_{x, y} \, \xi_y$ at the beginning of each Molecular Dynamics trajectory \cite{MontvayMuenster, DeGrandDeTarLQCD}.
   Nonzero mass term in (\ref{lat_act_interact}) and (\ref{det_str}) is necessary in order to ensure the invertibility of the staggered Dirac operator. Numerical results in the physical
    limit of zero mass were obtained by performing simulations at several nonzero values of $m$ and by extrapolating the expectation values of physical observables to $m \rightarrow 0$.

 In order to speed up the simulations we also perform local heatbath updates of the gauge field outside of the graphene plane (at $x^3 \neq 0$) between Hybrid Monte-Carlo updates.
 Both algorithms satisfy the detailed balance condition for the weight (\ref{ferm_integrated}) \cite{MontvayMuenster, DeGrandDeTarLQCD}. Successive application of these algorithms does not,
  in general, have this property. Nevertheless, by using the composition rule for transition probabilities it is easy to demonstrate that the path integral weight (\ref{ferm_integrated})
  is still the stationary probability distribution for such a combination of both algorithms. While local heatbath updates are computationally very cheap, they significantly decrease the
  autocorrelation time of the algorithm.

\subsection{Physical observables on the lattice}
\label{subsec:observables}

 The main goal of this paper is to measure the electric conductivity of graphene, that is, a linear response of the electric current density $J_i\lr{x} = \bar{\psi}\lr{x} \, \gamma_i \, \psi\lr{x}$ to the applied homogeneous electric field $E_j\lr{t}$ (where $t$ is the real Minkowski time). It is convenient to introduce the AC conductivity $\sigma_{ij}\lr{w}$, so that $\tilde{J}_i\lr{w} = \sigma_{ij}\lr{w} \, \tilde{E}_j\lr{w}$, where $\tilde{J}_i\lr{w} = \int dt \, e^{-i w t} J_i\lr{t}$ and $\tilde{E}_j\lr{w} = \int dt \, e^{-i w t} E_j\lr{t}$. Due to rotational symmetry of the effective field theory (\ref{partfan}), $\sigma_{ij}\lr{w}$ should have the form $\sigma_{ij}\lr{w} = \delta_{ij} \, \sigma\lr{w}$. Correspondingly, the DC conductivity is equal to the value of $\sigma\lr{w}$ at $w \rightarrow 0$.

 By virtue of the Green-Kubo dispersion relations \cite{Kadanoff:63:1, Asakawa:01:1, Aarts:07:1}, the Euclidean current-current correlators
\begin{eqnarray}
\label{corr}
 G\lr{\tau} = \frac{1}{2} \, \sum\limits_{i=1,2} \, \int dx^1 \, dx^2 \, \langle J_i\lr{0} \, J_i\lr{x} \rangle
\end{eqnarray}
can be expressed in terms of $\sigma\lr{w}$ as
\begin{eqnarray}
\label{corr_eq}
 G\lr{\tau} = \int\limits^{\infty}_{0}
 \frac{dw}{2 \pi} \, K\lr{w, \tau} \, \sigma\lr{w} ,
\end{eqnarray}
where the thermal kernel $K\lr{w, \tau}$ is
\begin{eqnarray}
\label{kernel}
 K\lr{w, \tau} = \frac{w \cosh\lr{w \lr{\tau - \frac{1}{2T}} }}{\sinh\lr{\frac{w}{2T}}}
\end{eqnarray}
and $\tau \equiv x^0$ is the Euclidean time. We use here a nonstandard definition of the kernel (\ref{kernel}) from \cite{Aarts:07:1}, which is more convenient for numerical analysis.

 Note that the current density in graphene is the charge which flows through the unit length in unit time and thus has the dimensionality of $L^{-2}$ (where $L$ stands for length) in units with $\hbar = c = 1$. Correspondingly, the current density in lattice units is the charge which flows through a link of the dual lattice of length $a$ in time $a/v_F$. Thus in order to express the current-current correlator (\ref{corr}) in physical units, one should multiply the result obtained on the lattice by $a^2 \, v_F^2/a^4$, where an additional factor of $a^2$ comes from integration over $x^1$, $x^2$ in (\ref{corr}). With the Euclidean time $\tau$ in (\ref{corr}), (\ref{corr_eq}) and (\ref{kernel}) being expressed in units of lattice spacing in temporal direction $a/v_F$, integration over $w$ in (\ref{corr_eq}) also includes a factor $v_F^2/a^2$. We thus conclude that the AC conductivity $\sigma\lr{w}$ is dimensionless. Moreover, the DC conductivity $\sigma\lr{0}$ is a universal quantity which does not depend on the lattice spacing or on the ratio of lattice spacings in temporal and spatial directions. For conversion to the SI system of units, it should be multiplied by $e^2/\lr{2 \pi h}$.

 In numerical simulations $G\lr{\tau}$ is measured for several ($\sim 10^1$) discrete values of $\tau$. A commonly used method to invert the relation (\ref{corr_eq})
  and to extract the continuum function $\sigma\lr{w}$ from the lattice discretization of $G\lr{\tau}$ is the Maximum Entropy Method \cite{Asakawa:01:1, Aarts:07:1}.

 For staggered fermions the electric current $J_i\lr{y}$ can be expressed in terms of the fields $\Psi_x$ as \cite{DeGrandDeTarLQCD}:
\begin{eqnarray}
\label{j_staggered}
 J_i\lr{y}  =
 \frac{1}{8} \sum\limits_{\eta} \,
 \delta_{\eta_3, \, 0} \, \delta_{\eta_i, \, 0} \,
 \left(
 \bar{\Psi}_{2 y + \eta} \alpha_{\eta, \, i} \Psi_{2 y + \eta + \hat{i}}
 + \right. \nonumber \\ \left. +
 \bar{\Psi}_{2 y + \eta + \hat{i}}
 \alpha_{\eta, \, i} \Psi_{2 y + \eta}
 \right)  ,
\end{eqnarray}
where we have taken into account that the spatial link variables $\theta_{x, i}$ are effectively equal to zero. Since the current (\ref{j_staggered}) is defined on the lattice
 with double lattice spacing, we calculate the Euclidean current-current correlator (\ref{corr}) only on time slices with even $\tau$.

 In order to make sure that we reproduce the results of \cite{Lahde:09:1, Lahde:09:2, Lahde:09:3, Drut:10:1, Lahde:11:1}, we have also calculated the fermionic chiral condensate.
  In terms of staggered fermions it can be written as
\begin{eqnarray}
\label{condensate_def}
 \vev{ \bar{\psi} \, \psi }
 =
 \frac{1}{8 \, L_0 \, L_1 \, L_2} \, \sum\limits_{x} \, \vev{\bar{\Psi}_x \Psi_x }  .
\end{eqnarray}

 After the fermions in the partition function are integrated out, the current-current correlator (\ref{corr}) and the chiral condensate (\ref{condensate_def})
  can be expressed in terms of expectation values of certain combinations of the staggered fermion propagator $D^{-1}_{x, y}\lrs{\theta_{x, \mu}}$ with respect to the weight
   (\ref{ferm_integrated}). We give the explicit expressions for these combinations in Appendix \ref{appsec:ferm_obs}.

\section{Numerical results}
\label{sec:num_results}

 Using the algorithm described in Subsection \ref{subsec:algorithms}, we have generated $400$ statistically independent gauge field configurations on the $20^4$ lattice for each point in the space of lattice parameters $\beta$ and $m$. For each value of $\beta$ in the range $\beta = 0.05 \ldots 0.025$ (which corresponds to substrate dielectric permittivities $\epsilon = 1.75 \ldots 12.75$ according to (\ref{lattice_coupling_constant})) the measurements were performed at three different
values of mass $m = 0.01, \, 0.02, \, 0.03$. For the smallest mass $m = 0.005$, for which the simulations are most expensive, $\beta$ took values in the range $\beta = 0.05 \ldots 0.15$  ($\epsilon = 1.75 \ldots 7.25$). In order to estimate the finite-volume effects, we have also generated $100$ gauge field configurations on the $28^4$ lattice with $\beta = 0.05 \ldots 0.21$ and $m = 0.01$.

\begin{figure}[ht]
 \includegraphics[width = 8.5cm]{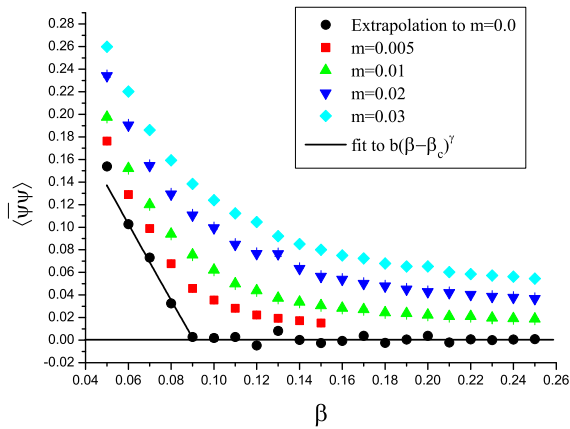}\\
 \caption{Fermionic condensate $\vev{\bar{\psi} \, \psi}$ as a function of inverse lattice coupling constants $\beta$ at different values of mass $m$
 and extrapolation to the limit $m \rightarrow 0$.
 Solid line is the fit of the extrapolated data with the function
 $\vev{\bar{\psi} \, \psi} \sim \lr{\beta_c - \beta}^\gamma$ with $\beta_c = 0.0908 \pm 0.0018$ and $\gamma = 1.0 \pm 0.16$.}
 \label{fig:condensates}
\end{figure}

 To check our simulation algorithms and to make sure that we reproduce the results of \cite{Lahde:09:1, Lahde:09:2}, we first consider the fermionic chiral condensate   $\vev{\bar{\psi} \, \psi}$. It is plotted on Fig.~\ref{fig:condensates} for different values of the mass $m$. The condensate rapidly decreases as the inverse lattice coupling constant $\beta$ (or, equivalently, the substrate dielectric permittivity $\epsilon$ (\ref{lattice_coupling_constant})) is increased. For each value of $\beta$,    we have fitted the mass dependence of the condensate with a quadratic polynomial and used this fit to extrapolate the data to the limit $m \rightarrow 0$. The result of such extrapolation is also shown on Fig.~\ref{fig:condensates}. It suggests that there is a critical value of $\beta$ in the range $0.08 < \beta_c < 0.09$ (which corresponds to $3.4 < \epsilon_c < 4.0$) such that the condensate is zero for $\beta > \beta_c$. A fit of the extrapolated data of the form $\vev{\bar{\psi} \, \psi} = b \lr{\beta_c - \beta}^\gamma$ for $\beta < \beta_c$ yields $\beta_c = 0.091 \pm 0.002$($\epsilon_c = 4.0 \pm 0.1$) and $\gamma=1.0 \pm 0.16$. These values are in agreement with the results of \cite{Lahde:09:1, Lahde:09:2}, where the critical inverse coupling constant        $\beta_c$ and the critical index $\gamma$ were estimated as $0.071 \geq \beta_c \leq 0.091$ and $\gamma \backsimeq 1$.

\begin{figure}[ht]
 \includegraphics[width=8.5cm]{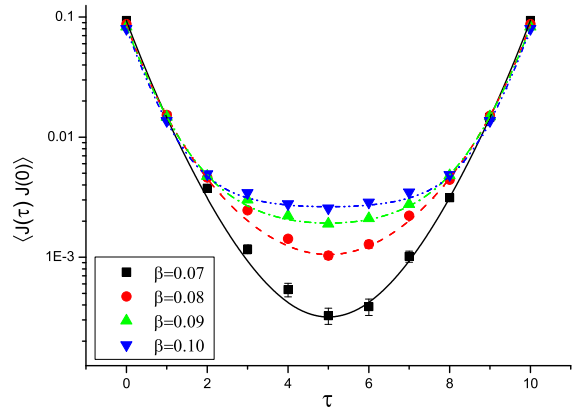}\\
 \caption{Current - current correlators (\ref{corr}) for $m = 0.01$. Solid lines are the fits obtained using the Maximum Entropy Method.}
 \label{fig:jj1}
\end{figure}

\begin{figure*}[ht]
 \includegraphics[width=8cm]{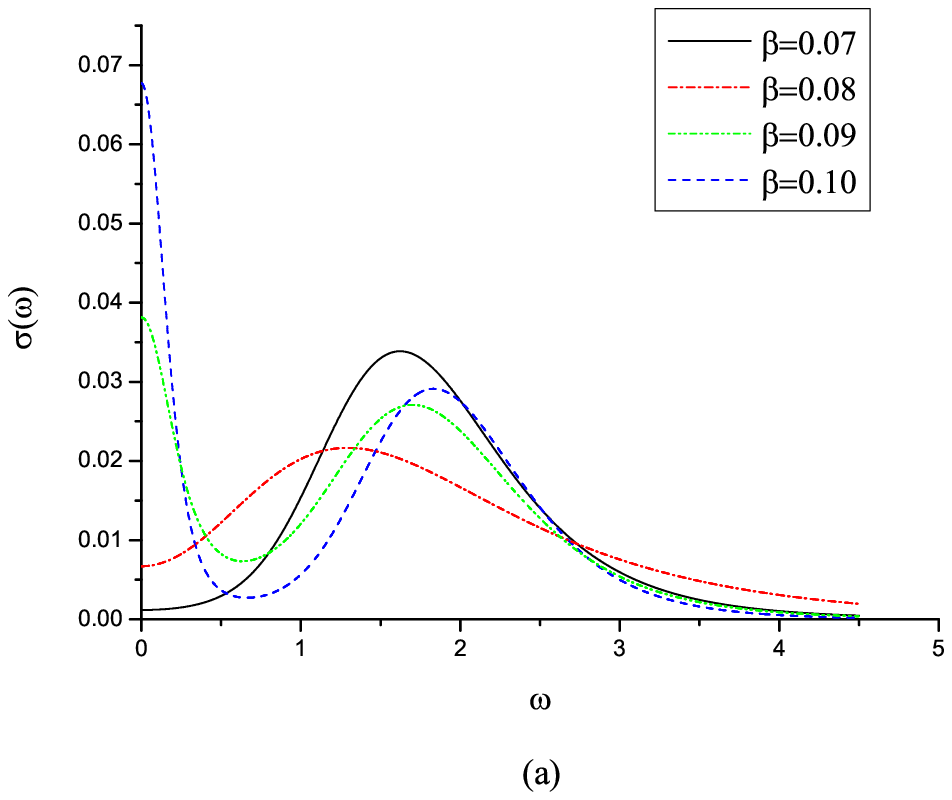}
 \includegraphics[width=8cm]{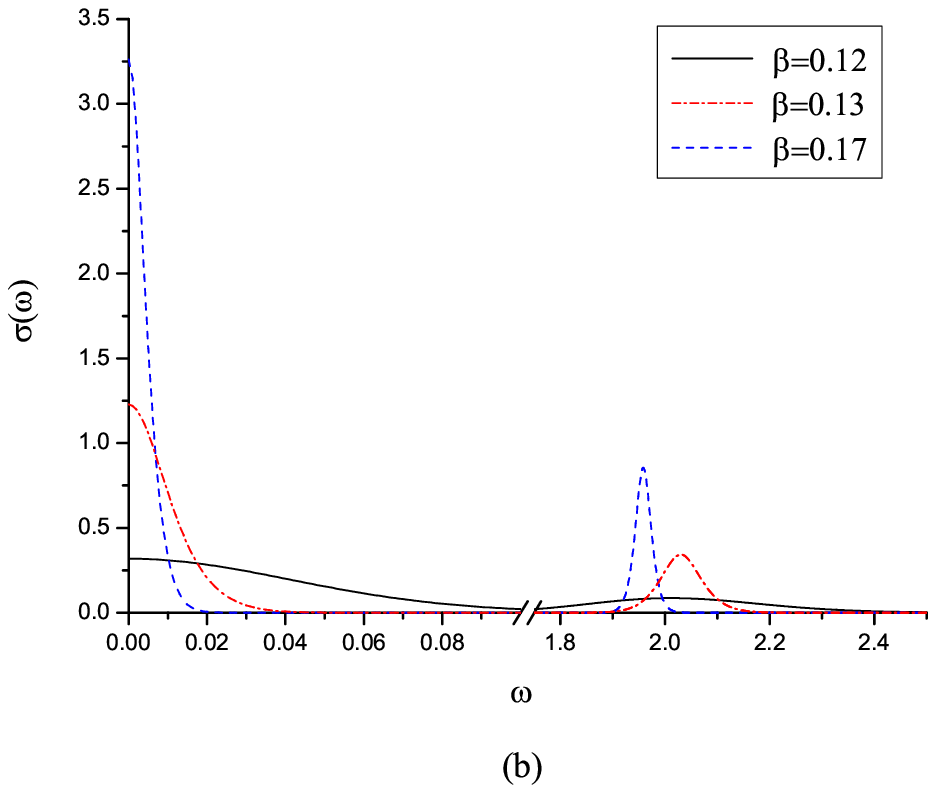}\\
 \caption{AC conductivity $\sigma\lr{w}$ in (\ref{corr_eq}) for $m=0.01$ at $\beta$ close to $\beta_c$ (on the left) and at $\beta > \beta_c$ (on the right).}
 \label{fig:rho}
\end{figure*}

 Our measurements of the conductivity of graphene start from the calculation of the current-current correlators (\ref{corr}), which are shown on Fig.~\ref{fig:jj1} for $m = 0.01$ and for different values of inverse coupling constant $\beta$. We note that the correlators decay significantly faster for smaller values of $\beta$.

 The AC conductivity $\sigma\lr{w}$ is extracted from the correlators using the Maximum Entropy Method with four basis eigenfunctions of the kernel (\ref{kernel}) and a constant model function $\sigma_0\lr{w} = 0.1$ \cite{Asakawa:01:1, Aarts:07:1}. The profiles of $\sigma\lr{w}$ are plotted on Fig.~\ref{fig:rho} and the corresponding fits of the correlators are shown on Fig.~\ref{fig:jj1} with solid lines.

 It is important to note at this point that on Fig.~\ref{fig:rho} the angular frequency $w$ is given in lattice units. For qualitative comparison with experimental data one can assume that the lattice spacing $a$ for the spatial directions of the cubic lattice used in our simulations is comparable with the lattice spacing $a = 0.246 \, {\rm nm}$ of the hexagonal lattice in graphene. After the rescaling (\ref{rescaling}) the discretization step for the Euclidean time $\tau$ should be of order ${\Delta \tau} \sim a/v_F$. The temperature $T$ in (\ref{corr_eq}) and (\ref{kernel}) is then equal to $k T = \hbar/\lr{L_0 \, {\Delta \tau} } \sim 0.1 \, {\rm eV}$, which is much smaller than the characteristic binding energy in graphene $\sim 1 \, {\rm eV}$. Thus our simulation results should correspond to sufficiently low physical temperatures as compared to characteristic excitation energies.

 For the inverse coupling constant $\beta$ below approximately $0.08$ ($\epsilon<3.4$), $\sigma\lr{w}$ has one very broad peak around $w \approx 1.2$, and the DC conductivity $\sigma\lr{0}$ has some small nonzero value. As $\beta$ increases towards $\beta_c$, the second peak emerges at $w = 0$ and both peaks become narrower and higher (see Fig.~\ref{fig:rho}, left plot). The emergence of the second peak results in the rapid growth of the DC conductivity. At $\beta > \beta_c$, the two peaks continue to grow, and their widths become comparable to the temperature $T = L_0^{-1}$ in lattice units (see Fig.~\ref{fig:rho}, right plot).

 In order to understand such peak structure in the weak-coupling limit, remember that for free Dirac fermions the AC conductivity $\sigma\lr{w}$ has a delta-function singularity at $w = 0$ \cite{Katsnelson:06:1, Miransky:02:1}, which is a manifestation of the absence of scattering of charge carriers. When the interactions are turned on, this peak is smeared, which results in a large but finite value of the DC conductivity $\sigma\lr{0}$. The second peak practically does not move as the mass $m$ is changed. We conjecture that this second peak corresponds to a saddle point in the dispersion relation of staggered fermions which is situated in the middle of a straight line which connects the two distinct Fermi points \cite{Buividovich:12:1}. The position of this peak should thus depend on the lattice regularization of the effective field theory (\ref{partfan}) and should correspond to the optical frequency range for real graphene.

\begin{figure}[ht]
 \includegraphics[width=8.5cm]{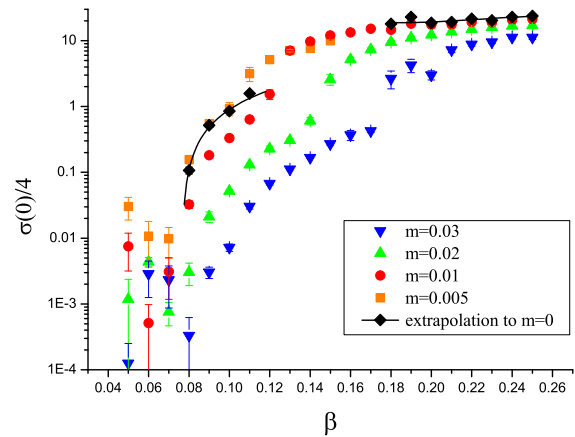}\\
 \caption{DC conductivity per spin per valley $\sigma\lr{0}/4$ in units of $e^2/h$ as a function of inverse lattice coupling constants at different values of mass $m$. The result of extrapolation of the data to the limit $m \rightarrow 0$ is plotted with black dots and solid lines.}
 \label{fig:conductivity}
\end{figure}

\begin{figure}[ht]
 \includegraphics[width=8cm]{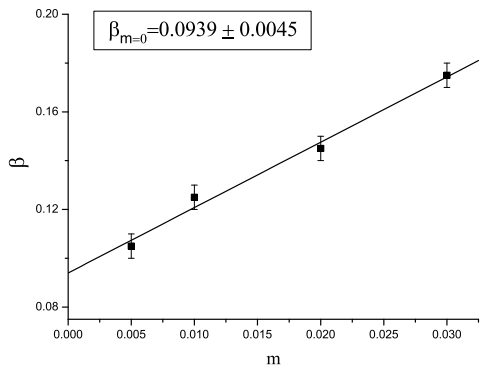}\\
 \caption{Linear extrapolation of the position of the second discontinuity of the DC conductivity to the limit  $m\rightarrow 0$.}
 \label{fig:FitTransition}
\end{figure}

 The DC conductivity $\sigma\lr{0}$ is shown on Fig.~\ref{fig:conductivity} as a function of inverse coupling constant $\beta$ at different values of the mass $m$. We normalize the conductivity to a single spin component and a single Dirac point, thus on Fig.~\ref{fig:conductivity} we plot $\sigma\lr{0}/4$ rather than $\sigma\lr{0}$. $\sigma\lr{0}$ quickly decreases as both $m$ and $\beta$ become smaller. One can also note two distinct discontinuities in the dependence of $\sigma\lr{0}$ on $\beta$. For example, at $m=0.01$ the first discontinuity is situated between $\beta=0.07$ ($\epsilon = 3$) and $\beta=0.09$ ($\epsilon = 4$), and the second one - between $\beta=0.12$ ($\epsilon=5.6$) and $\beta=0.13$ ($\epsilon=6.2$). The position of the first discontinuity depends only weakly on the mass $m$ and roughly corresponds to the critical inverse coupling constant $\beta_c$ obtained from the analysis of the chiral condensate (see Fig.~\ref{fig:condensates}). The second discontinuity shifts to smaller $\beta$ and becomes somewhat weaker as $m$ decreases. Linear extrapolation of the position of this discontinuity to the limit of zero mass (see Fig.~\ref{fig:FitTransition}) suggests that at $m = 0$ both discontinuities coincide. We also note that the profile of the AC conductivity $\sigma\lr{w}$ practically does not change across this second discontinuity. Thus there seems to be a single phase transition in the chiral limit $m \rightarrow 0$, in agreement with the results of \cite{Lahde:09:1, Lahde:09:2, Lahde:09:3}. The corresponding critical value of the inverse coupling constant can be estimated to lie in the range $0.07 \le \beta_c \le 0.09$ ($3 \le \epsilon_c \le 4$).

 For each value of $\beta$ between the two discontinuities, we also perform the quadratic fit of the mass dependence of the conductivity and use it to extrapolate the data to $m \rightarrow 0$. This extrapolation is shown on Fig.~\ref{fig:conductivity} with black dots and solid lines.

\begin{figure}[ht]
 \includegraphics[width=8.5cm]{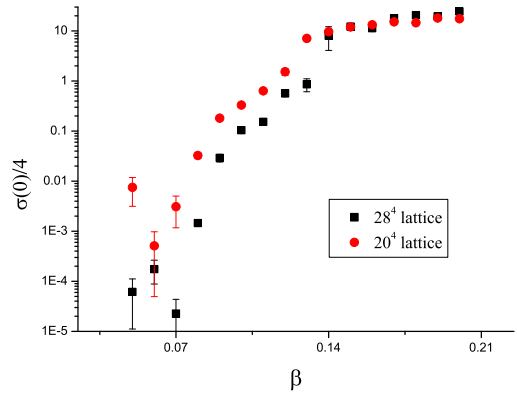}\\
 \caption{DC conductivity per spin per valley $\sigma\lr{0}/4$ in units of $e^2/h$ as a function of inverse lattice coupling constant $\beta$ for $20^4$ and $28^4$ lattices with $m=0.01$.}
 \label{fig:conductivity28}
\end{figure}

 On Fig.~\ref{fig:conductivity28} we compare the DC conductivity at $m = 0.01$ on $20^4$ and $28^4$ lattices. The dependence of $\sigma\lr{0}$ on the inverse coupling constant $\beta$ is qualitatively the same for both lattices, in particular, the positions of the discontinuities practically coincide. However, the actual values of the DC conductivity differ beyond the error bars, especially in the strong-coupling phase. This suggests that finite-volume and finite-temperature effects could be quite large for our lattice parameters. Indeed, quite large finite-temperature effects have been reported in a recent Monte-Carlo study of the tight-binding model on the hexagonal lattice \cite{Buividovich:12:1}, where the values of lattice parameters were quite close to those used in this work. We leave the detailed study of finite-temperature and finite-volume effects as a direction for further investigations.

\section{Conclusions}
\label{sec:conclusions}

 In this paper we have numerically studied the AC and DC conductivities of graphene by using lattice Monte-Carlo simulations with $2 + 1$-dimensional staggered fermions which interact with $3 + 1$-dimensional non-compact Abelian lattice gauge field. We have found that in a phase with spontaneously broken chiral symmetry (which corresponds to sublattice symmetry of the original hexagonal lattice) the DC conductivity rapidly decreases as the substrate dielectric permittivity becomes smaller. The estimates of the corresponding critical values $0.07 \le \beta_c \le 0.09$ ($3 \le \epsilon_c \le 4$) obtained both from the measurements of the chiral condensate and the conductivity agree with each other and with the results of \cite{Lahde:09:1, Lahde:09:2, Lahde:09:3, Lahde:11:1}. This supports the existence of a single insulator-semimetal phase  transition in graphene. Interestingly, our estimate of $\epsilon_c$ is close to the dielectric permittivity of silicon dioxide $\epsilon_{SiO_2} = 3.9$, which is often used as a substrate for graphene. According to the data presented on Fig.~\ref{fig:conductivity}, for the largest value of the coupling constant used in our simulations ($\beta = 1/g^2 = 0.05$) the DC conductivity turns out to be smaller than the DC conductivity in the semimetal phase (at $\beta > \beta_c$) by a factor of order of $10^3$.

 Finally, we note that our value of the DC conductivity in the semimetal phase is significantly larger than the conductivity of non-interacting quasiparticles in the ideal monolayer graphene ($\sigma\lr{0} \sim 4 e^2/h$) obtained in \cite{Vildanov:09:1, Katsnelson:06:1, Novoselov:09:1} using the Landauer approach. However, as was stressed in \cite{Katsnelson:06:1}, in Landauer approach a finite value of the conductivity of free Dirac fermions is determined solely by scattering on the boundaries of the sample. In the absence of boundaries (for instance, on the lattice with torus topology) the AC conductivity $\sigma\lr{w}$ has a delta-function singularity at $w = 0$, and thus the DC conductivity $\sigma\lr{0}$ is formally infinite. In an interacting theory, this singularity is smeared out, and the DC conductivity takes some finite (but large) value.

\begin{acknowledgments}
 The authors are much obliged to Dr. Timo Lahde who was the first to draw their attention to  graphene field theory. The authors are grateful to Prof. Mikhail Zubkov for interesting and useful discussions. The work was supported by Grant RFBR-11-02-01227-a and by the Russian Ministry of Science and Education, under contract No. 07.514.12.4028. Numerical calculations were performed at the  ITEP
system Stakan (authors are much obliged to A.V. Barylov, A.A. Golubev, V.A. Kolosov, I.E. Korolko, M.M. Sokolov for the help), the MVS 100K at Moscow Joint Supercomputer Center and at Supercomputing Center of the Moscow State University.
\end{acknowledgments}

\appendix
\section{Calculation of fermionic observables}
\label{appsec:ferm_obs}

 Here we give explicit expressions for the vacuum expectation values of fermionic observables used in our simulations.

 Fermionic chiral condensate corresponds to the diagonal elements of the staggered fermion propagator:
\begin{eqnarray}
 \vev{\bar{\psi} \, \psi}
 =
 \frac{1}{8 \, L_0 \, L_1 \, L_2} \, \sum\limits_{x} \, \vev{D^{-1}_{x, x} }
\end{eqnarray}
where $\vev{\ldots}$ on the right-hand side denotes averaging over lattice gauge field $\theta_{x, \mu}$ with the weight (\ref{ferm_integrated}).
 We calculate this trace using the stochastic estimator \cite{MontvayMuenster, DeGrandDeTarLQCD}.

 Current-current correlator is a sum of connected and disconnected parts:
\begin{equation}
\label{conn_disconn}
 G\lr{y^0} = C\lr{y^0} - D\lr{y^0}  .
\end{equation}
The connected contributions can be expressed in terms of staggered fermion propagator as \cite{DeGrandDeTarLQCD}:
\begin{eqnarray}
\label{conn}
 C\lr{y^0} = \frac{1}{64} \, \sum\limits_{y_1, y_2}
 \sum\limits_{\eta, \eta^{\prime}}
 \alpha_{\eta, i} \alpha_{\eta^{\prime}, i}
 \times \nonumber \\ \times
 \left(  \vev{
  S\lr{2 y + \eta, \hat{i} + \eta^{\prime}}
  S\lr{\eta^{\prime}, 2 y + \hat{i} + \eta}
 }
 \right. + \nonumber \\ +
  \vev{
   S\lr{2 y + \eta, \eta^{\prime}}
   S\lr{\hat{i} + \eta^{\prime}, 2 y + \hat{i} + \eta}
  }
 + \nonumber \\ +
  \vev{
   S\lr{2 y + \hat{i} + \eta, \hat{i} + \eta^{\prime}}
   S\lr{\eta^{\prime}, 2 y + \eta}
  }
 + \nonumber\\ + \left.
 \vev{
  S\lr{2 y + \hat{i} + \eta, \eta^{\prime}}
  S\lr{\hat{i} + \eta^{\prime}, 2 y + \eta}
  } \right)  ,
\end{eqnarray}
where $S\lr{x, y} \equiv D^{-1}_{x, y}$.

 For the calculation of the connected part of the correlator (\ref{conn_disconn}) we take into account that the solution of the linear equation $\chi_y = D_{y, z} \psi_z$
  with $\chi_y = \delta_{x, y}$ yields the staggered fermion propagator $D^{-1}_{x, y}$ for all $y$.

 The disconnected part takes the following form:
\begin{eqnarray}
\label{disconn}
 D\lr{y^0} =
 \frac{1}{64} \sum\limits_{y_1, y_2}
 \sum\limits_{\eta, \eta^{\prime}}
 \alpha_{\eta, i} \, \alpha_{\eta^{\prime}, i}
 \nonumber \\
 \langle \lr{
  S\lr{\eta^{\prime}, \hat{i} + \eta^{\prime}}
  +
  S\lr{\hat{i} + \eta^{\prime}, \eta^{\prime}}
 }
 \times \nonumber\\ \times
 \lr{
  S\lr{2 y + \hat{i}, 2 y + \eta}
  +
  S\lr{2 y + \eta, 2 y + \hat{i} + \eta}
 }
 \rangle  .
\end{eqnarray}

 In practice the disconnected part of the correlator (\ref{conn_disconn}) is calculated using stochastic estimators \cite{MontvayMuenster, DeGrandDeTarLQCD},
  similarly to the chiral condensate. In our simulations we have found that the disconnected part of the correlator is much smaller and much noisier than the
   connected one. Therefore we have neglected it in our measurements of the conductivity.


\end{document}